\documentstyle[preprint,aps,epsfig]{revtex}

\begin{document}
\draft
\title{Sinai billiards, Ruelle zeta-functions and Ruelle resonances:  
microwave experiments}
\author{S. Sridhar and W. T. Lu}
\address{Department of Physics\\
Northeastern University, Boston, MA 02115}
\maketitle

\begin{abstract}
We discuss the impact of recent developments in the theory 
of chaotic dynamical 
systems, particularly the results of Sinai and Ruelle, 
on microwave experiments 
designed to study quantum chaos. The properties of 
closed Sinai billiard microwave
cavities are discussed in terms of universal 
predictions from random matrix theory, 
as well as periodic orbit contributions which
manifest as `scars' in eigenfunctions. 
The semiclassical and classical Ruelle
zeta-functions lead to quantum and 
classical resonances, both of which are
observed in microwave experiments on 
$n$-disk hyperbolic billiards.
\end{abstract}

\medskip
\noindent
{\bf key words}: microwave, hyperbolic, Sinai billiard, 
correlation, Ruelle zeta-function,
resonances

\medskip
\pacs{}

\section{Introduction}

It may come as a pleasant surprise to the dynamical systems community, to
learn that the work of Yasha Sinai and David Ruelle has had a major impact
in experiments on microwave geometries. The connections arise from recent
developments which demonstrate that wave mechanics experiments using
microwaves are an ideal laboratory for studying the so-called
quantum-classical correspondence, a central issue in quantum chaos.
These microwave experiments have shown that several theoretical results on
the mathematics of chaotic dynamical systems have manifestations in their
corresponding quantum or wave mechanics.

The term chaos often means hyperbolicity which guarantees the decomposition
of the tangent space at each phase space point into expansion and
contraction subspaces. A systematic study of hyperbolic systems from the
geometric theory perspective was initiated by Smale \cite{Smale}. Another
viewpoint, namely the ergodic theory or the probability approach, was
pioneered by Sinai and Ruelle. Of great interest and possible experimental
realization are billiard systems which are two-dimensional, bounded or open.
Not all closed billiards have hyperbolic properties. For a dispersing billiard
which is closed and has finite concave boundaries, Sinai \cite{Sinai} proved
that the system is ergodic. But certain billiards with convex boundaries 
\cite{Wojtkowski} such as the Bunimovich stadium \cite{Bunimovich}, were
also found to be hyperbolic. Another important billiard system is 
that of $n$ disks on a plane,
which is hyperbolic and a paradigm model of Axiom A systems
considered by Ruelle \cite{Ruelle} and Pollicott \cite{Pollicott}.

Experiments utilizing thin 2-D microwave structures have been shown recently
to provide an ideal laboratory system to explore issues in hyperbolic
dynamics and quantum chaos. These experiments utilize thin geometries in
which Maxwell's equations reduce to the time-independent Schr\"{o}dinger
equation $(\nabla ^{2}+k^{2})\Psi =0$, where $\Psi $ is $E_{z}$ the $z$%
-component of the electric field. This mapping enables precision room
temperature experiments exploring wave mechanics on laboratory length scales
that can be easily manipulated. The experiments typically probe the
two-point Green's function of geometries which may be {\em closed}, such as
the Sinai billiard, or {\em open}, such as $n$ disks on a plane. In addition
to yielding eigenvalue or resonance spectra, a unique advantage of the
microwave experiments is the ability to directly map eigenfunctions and
standing wave patterns.

A major theme of work in this area has focussed on the quantum-classical
boundary. Some of the principal results that have emerged are the
observation of scars \cite{Sridhar91}, precision tests of random matrix
theory (RMT) and non-linear sigma models in chaotic and disordered cavities 
\cite{Kudrolli94,Pradhan00}, observation of localization in disordered
billiards \cite{Kudrolli95}, and classical and quantum properties of open
systems \cite{Lu99,Pance00}.

These microwave experiments may also be classified as {\it experimental
mathematics}! A noteworthy example is the work on ``Not `Hearing The Shape'
of Drums'' \cite{Sridhar94}, which demonstrated experimentally the exact
equivalence of eigenvalues of certain isospectral geometries. The
isospectrality is purely a consequence of geometry or combinatorics,
and is independent of the dynamics (whether chaotic or regular).

This paper summarizes recent results exemplifying the impact of Sinai and
Ruelle's work on the microwave quantum chaos experiments. We first discuss
the properties of the eigenvalues and eigenfunctions of Sinai
billiard-shaped microwave cavities. We then discuss, in the context of 2-D $n
$-disk billiards, the major role of the Ruelle dynamical zeta-functions in
the classical and semiclassical theory of chaotic repellers, and the
experimental observation of Ruelle-Pollicott resonances.

\section{Experiments on Sinai billiards}

Following the pioneering work of Sinai, the Sinai billiard has become a
model geometry exemplifying classical and quantum\ chaos in a closed system. An experimental
realization of the Sinai billiard was first achieved in microwave cavities 
\cite{Sridhar91}.

The microwave experiments on Sinai billiard cavities yield the eigenvalue spectra $\{E_{n}=f_{n}^{2}\}$ directly
as peaks at frequencies $\{f_{n}\}$ in the experimental 2-point microwave
transmission function $S_{21}(f)$ of the cavity (Fig.1(top)).
Typically $\sim 600-1000$ eigenvalues can be measured in the experimentally
accessible frequency window. The corresponding stick spectrum is then
analyzed to yield information on spectral statistics. The approach here
closely parallels work in spectra of atomic and nuclear 
systems \cite{RMT98},
which have been analyzed in terms of RMT. Here we summarize the principal
results for the Sinai billiard. The main conclusion is that the eigenvalue
and eigenfunction statistics are in good agreement with universal
predictions of RMT expected to apply to quantum systems with classically
chaotic dynamics \cite{Stockmann}. However, sufficiently precise experiments
clearly reveal deviations from the universal results due to the presence of
marginal unstable ``bouncing ball'' (b-b) orbits.

The nearest neighbor spacing statistics $P(s)$ for the Sinai billiard
spectrum shows the well-known level repulsion $P(s)\rightarrow 0$ as the
spacing $s\rightarrow 0$, a characteristic feature of quantum chaos
(Fig.1(middle)). Fits to the Brody function $P(s)=As^{b}\exp (-Bs^{b+1})$, yield
a Brody parameter $b=0.92$, where $b=1.0$ corresponds to exact agreement
with the famous Wigner-Dyson statistic, $P(s)={\pi s\over 2}\exp (-{\pi \over 4}s^{2})$. This
deviation is a small but definite signature of the b-b orbits. More striking
manifestations of these orbits can be found in longer range correlations,
such as the spectral rigidity $\Delta _{3}(L)$ (Fig.1(bottom)). This quantity for
the Sinai billiard shows initial agreement with RMT for small $L$ followed
by an essentially linear rise from the weak logarithmic dependence expected.
Incorporation of the b-b orbits contributions gives excellent agreement
between experimental data and theory.

The eigenfunctions corresponding to each eigenvalue peak are obtained using
cavity perturbation techniques described in Ref. \cite{Sridhar92}.
Representative eigenfunctions of the Sinai billiard are shown in Fig.2, and
show a fascinating variety of patterns. This is in contrast with the
rectangular cavity, obtained from the Sinai billiard by removing the central
circular disk, for which each eigenfunction can be labeled by $\{m,n\}$,
which arise from the integrability of $\{k_{x},k_{y}\}$. Non-integrability
in the Sinai billiard is immediately apparent from the fact that the only
quantum number that can be assigned to each eigenfunction is $E_{n}$ - the
lack of integrals of motion precludes any further quantum number
assignments. The question, what classical phase space structures
characterize the eigenfunctions, has occupied physicists for several years.
Heller pointed out that the eigenfunctions of chaotic billiards should
organize themselves around periodic orbits, even the unstable ones, and
called these structures ``scars''\cite{Heller}. Scars were seen in numerical
simulations, and were implicated indirectly in atomic phenomena. The first
direct experimental observation of scars was in eigenfunctions of a Sinai
billiard microwave cavity\cite{Sridhar91}. An example is shown in Fig.2 at $%
f=3663MHz$. This eigenfunction is scarred along the diagonal periodic orbit,
which is unstable. The presence of stable b-b orbits is also visible in some
eigenfunctions such as the one at $f=2866MHz$. However most of the
eigenfunctions are not {\em visibly }scarred by either stable or unstable
PO. A favorite is the one at $f=3767MHz$. It shows the circular symmetry of
the central disk with almost no visible influence of the rectangular
boundary. A more complete characterization of the eigenfunctions in terms of
classical phase space structures is a major challenge of research in this
area.

Another approach to analyzing eigenfunctions is in terms of their
statistical properties. The leading statistic is the density distribution $%
P(|\Psi |^{2})$ which is shown in Fig.3(top). For the quarter Sinai billiard,
in which the geometric symmetries are absent, the finite probability of
finding large densities is evident from the figure. The dependence of $%
P(|\Psi |^{2})$ on the density $|\Psi |^{2}$ follows closely, but not
exactly, the Porter-Thomas (P-T) distribution obtained from RMT. This again
is due to the presence of the b-b orbits. We have confirmed this by showing
almost exact agreement with the P-T distribution by removing the b-b orbits
by constructing a so-called Sinai stadium, which is a hybrid of the Sinai
and stadium billiards.

Spatial correlations $\left\langle |\Psi ({\bf r})|^{2}|\Psi ({\bf r}%
^{\prime })|^{2}\right\rangle $ can also be examined and are shown in
Fig.3(bottom). The experimental data follow closely the dependence $\left\langle
|\Psi ({\bf r})|^{2}|\Psi ({\bf r}^{\prime })|^{2}\right\rangle
=1+cJ_{0}^{2}(k|{\bf r}-{\bf r}^{\prime }|)$, with $c=2$, expected from RMT
for 2-D. The data for the Sinai billiard lie just below this functional
form, this can be seen from the inverse participation ratio $\left\langle
|\Psi ({\bf r})|^{4}\right\rangle =2.7\pm 0.1$, obtained from the spatial
correlation with ${\bf r}={\bf r}^{\prime }$. The deviation from the
universal value of $3.0$ on the lower side is due to the b-b orbits. Here
again better agreement with the RMT universal value and function is obtained
by eliminating the b-b orbits such as in the Sinai stadium.

In summary the Sinai billiard has been a model system exemplifying the
fundamental properties of a quantum chaotic system, with near perfect
agreement with RMT predictions for universal distributions of spectral 
and eigenfunction statistics, and where the deviations 
from universality are clearly
identifiable as arising from b-b orbits.

\section{Ruelle zeta-functions and Ruelle-Pollicott resonances in $n$-disk
chaotic billiards}

The system of $n$ disks on a plane \cite{Gaspard89} provides us with a
paradigm model system for classical and quantum chaos in open systems. The
system is hyperbolic and not ergodic. The microwave realization of the $n$%
-disk system is similar to that of the Sinai billiard. To make the system
open, microwave absorbers were placed 
around and far from the disks to have experimentally
ignorable reflection, and to provide a good realization of escape to infinity. 
Because of the $C_{nv}$ symmetry of the $n$-disk
system, one can probe the system in many different ways. Periodic orbit
based treatments have been successfully used both classically and
semi-classically. The Ruelle dynamical zeta-function, in both the classical
and semiclassical versions, has played a major role in studying the
properties of this system.

The semiclassical Ruelle zeta-function is

\[
\zeta _{{\rm sc},j}(-ik)=\prod_{p}
\left( 1-\frac{e^{i(kL_{p}+\pi \mu_{p}/2)}}{\sqrt{
\left| \Lambda _{p}\right| }\Lambda_{p}^{j}}\right)^{-1} 
\]
and the classical Ruelle zeta-function is 
\[
\zeta _{{\rm cl},\beta }(s)=\prod_{p}
\left( 1-\frac{e^{-sT_{p}}}{\left| \Lambda
_{p}\right| \Lambda_{p}^{\beta -1}}\right)^{-1}. 
\]
The product is over all classical primitive periodic orbits without
repetition. Here $L_{p}$ and $T_{p}$ are the length and period of periodic orbit $p$ with the Maslov
index $\mu _{p}$. $\Lambda _{p}$ is the bigger eigenvalue of the monodromy
stability matrix for billiard systems.

The semiclassical Ruelle zeta-function can be derived from the Gutzwiller
trace formula \cite{Gutzwiller90} of the Green's function in the
semiclassical limit $\hbar \rightarrow 0$ which is summed over all classical
periodic orbits. The finite widths of the quantum resonances arise in the $n$-disk system because the system is open and the strange repeller made of classical orbits can trap
waves only for finite time. The classical Ruelle zeta-function can be
derived from the trace of the resolvent $(s-{\cal L})^{-1}$ with ${\cal L}%
f\equiv \{H,f\}$ for Hamiltonian system. For integrable system, the poles of
the resolvent operator are purely imaginary, this gives the periodic motion
of the classical dynamics. For open or ergodic system, these poles may have
nonzero real part and some of them may be real. This gives rise to the correlation decay
of the classical dynamics \cite{Young}. These complex poles are the
Ruelle-Pollicott (R-P) resonances \cite{Ruelle,Pollicott}. If the poles are
isolated, the decay is exponential. In the case of a branch cut, the decay may
be algebraic. Contrary to the trace formula of the classical resolvent \cite
{Cvitanovic91} which is exact, the semiclassical trace formula of Gutzwiller
is not, and it is not always guaranteed that the poles of the semiclassical
Ruelle zeta-function are also the poles of 
the Green's function. But for hard wall
systems such as the billiards, the discrepancy was found to be very small 
\cite{Gaspard94,Cvitanovic01}. The index $j$ and $\beta $ comes in from the
decomposition of the trace formulas into Ruelle zeta-functions with $%
j=0,1,\cdots $ and $\beta =1,2,\cdots $. One may extrapolate $\beta $ from
integers to real numbers to get more information of the classical dynamics 
\cite{Gaspard98}. For the calculation of sharp resonances, only $j=0$ and $%
\beta =1$ will be needed in the Ruelle zeta-functions.

For the $n$-disk system, since the velocity of the particle is constant, $%
L_{p}=\upsilon T_{p}$, one can simply set $\upsilon =1$, and work in the $k$%
-space for both the classical and semiclassical Ruelle zeta-functions. If the
separation of the disks are big enough to have no tangent orbits, the
dynamics is continuously hyperbolic and stable. A symbolic dynamics of $n-1$
symbols with finite grammar exists for the system. The Euler product of the
Ruelle zeta-functions can be expanded as $\zeta
^{-1}=\prod_{p}(1-t_{p})=1-\sum_{f}t_{f}+\sum_{r}c_{r}$. Here $t_{f}$ is the
weight of fundamental orbits $f$, $c_{r}$ is the curvature which is the
weight difference between long orbit and combination of shorter ones. Only a
few primitive POs will be needed in the Ruelle zeta-function to give quite
accurate calculation of the poles because the curvature in the cycle
expansion will decay exponentially with increasing PO length \cite
{Cvitanovic89}. This further simplifies the task of calculating resonances.
For the 3-disk system, the number of fundamental orbits is just 2. Only 8
periodic orbits with period up to 4 were be needed in the Ruelle
zeta-functions to give very accurate estimate of the quantum and classical
resonances.

A typical trace of the transmission $T(k)=|S_{21}(k)|^{2}$ of the quantum
resonances is shown in Fig.4 for the 3-disk system in the fundamental
domain. The experimental trace was fitted well by the sum of Lorentzians

\[
T(k)=\sum_{n}\frac{b_{n}}{(k-s_{n})^{2}+s_{n}^{\prime 2}}. 
\]
Here $s_{n}+is_{n}^{\prime }$ are the semiclassical resonances obtained
through the semiclassical Ruelle zeta-function. The coupling $b_{n}$ depend
on the probes location and were chosen to optimize the fitting.

The similarity between the classical Ruelle zeta-function and the
semiclassical ones is not coincidental. This implies a deeper relation
between the quantum and the classical dynamics besides the correspondence
principle. There is increasing interest in understanding the nature of the
classical-quantum correspondence. The quantum 
dynamics is found to approach the
classical dynamics with increasing coarse-graining in the phase space for a
mixed phase space system \cite{Haake}. The autocovariance of the two-level
spacing and the correlation of the density of states are also found to be
related to the classical zeta-function in the study of the Riemann
zeta-function zeroes \cite{Bohigas}. For an open hyperbolic system, we have
shown that the correlation decay of the quantum system is determined by the
classical dynamics \cite{Pance00}. That means the classical R-P resonances
display their fingerprints in the quantum dynamics. We find that the
autocorrelation of the transmission $C(\kappa )=\left\langle T(k)T(k+\kappa
)\right\rangle $ are well-described as a sum of Lorentzians

\[
C(\kappa )=\sum_{i,\pm}\frac{c_{i}}{(\kappa \pm\gamma_{i}^{^{\prime
}})^{2}+\gamma_{i}^2}
\]
Here $\gamma _{i}\pm i\gamma _{i}^{^{\prime}}$ are the 
{\em classical} R-P resonances. The coefficients $c_{i}$ are related to the
eigenfunctions of the Perron-Frobenius operator. The experimental
observation of the R-P resonances is illustrated in Fig.5.

We are thus able to observe both the quantum and classical resonances in the
same microwave experiment! This exactly parallels the periodic orbit
treatment, in which the poles of the semiclassical Ruelle zeta-function
yields the quantum resonances while the classical Ruelle zeta-function
provides the classical resonances for the hyperbolic $n$-disk system. The
convergence of experiment and theory have been made possible by extensive
developments in the machinery of both approaches. The advantages of the
microwave experiments are the well-specified nature of the hard wall
potentials, and the efficient ensemble averaging that can be achieved. The
microwave experiments thus have emerged as an ideal laboratory system to
study both quantum and classical chaos, and some of the mysteries of the
quantum-classical correspondence.

In closing, we express our appreciation of the influential role played by
the Statistical Physical Workshops at Rutgers University organized by
Professor Joel Lebowitz. Our observation of scars in Sinai billiard cavities
was first reported at the 1991 meeting, and the observation of R-P
resonances was reported in the 2000 meeting.

This work was supported by NSF-PHY-0098801.

\newpage
\begin{figure}[htbp]
\center{
\epsfig{figure=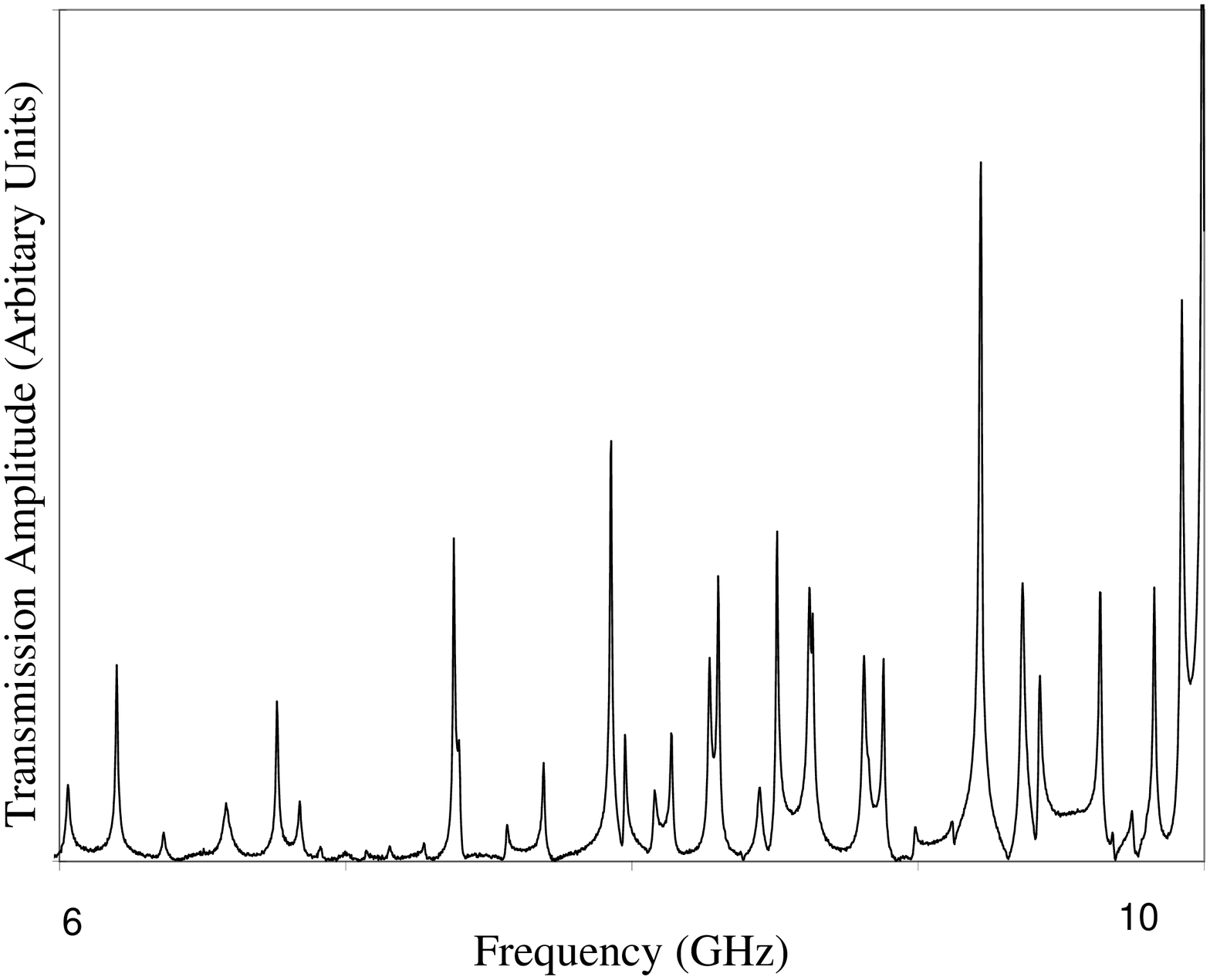,width=3.5in}
\epsfig{figure=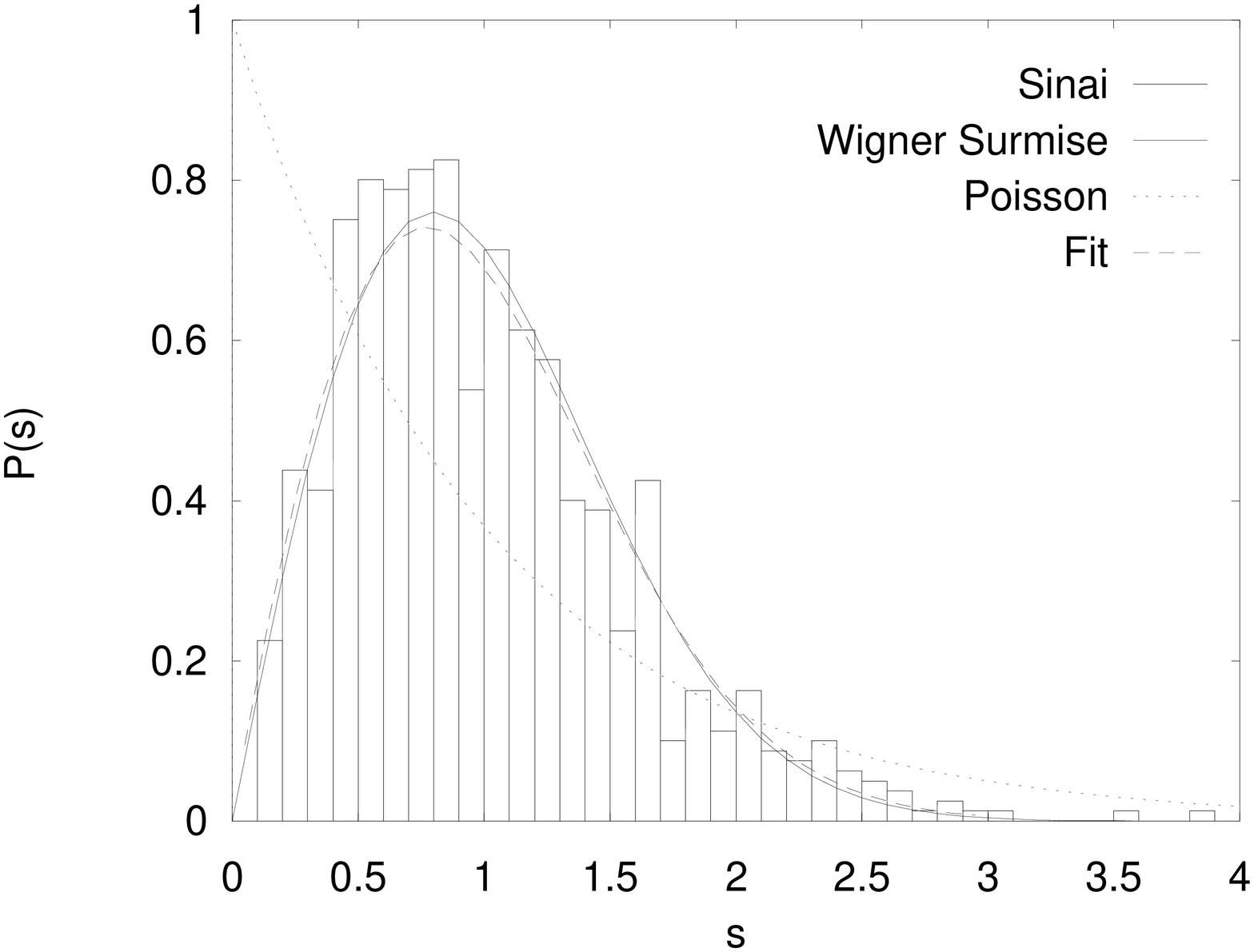,width=3.5in}
\epsfig{figure=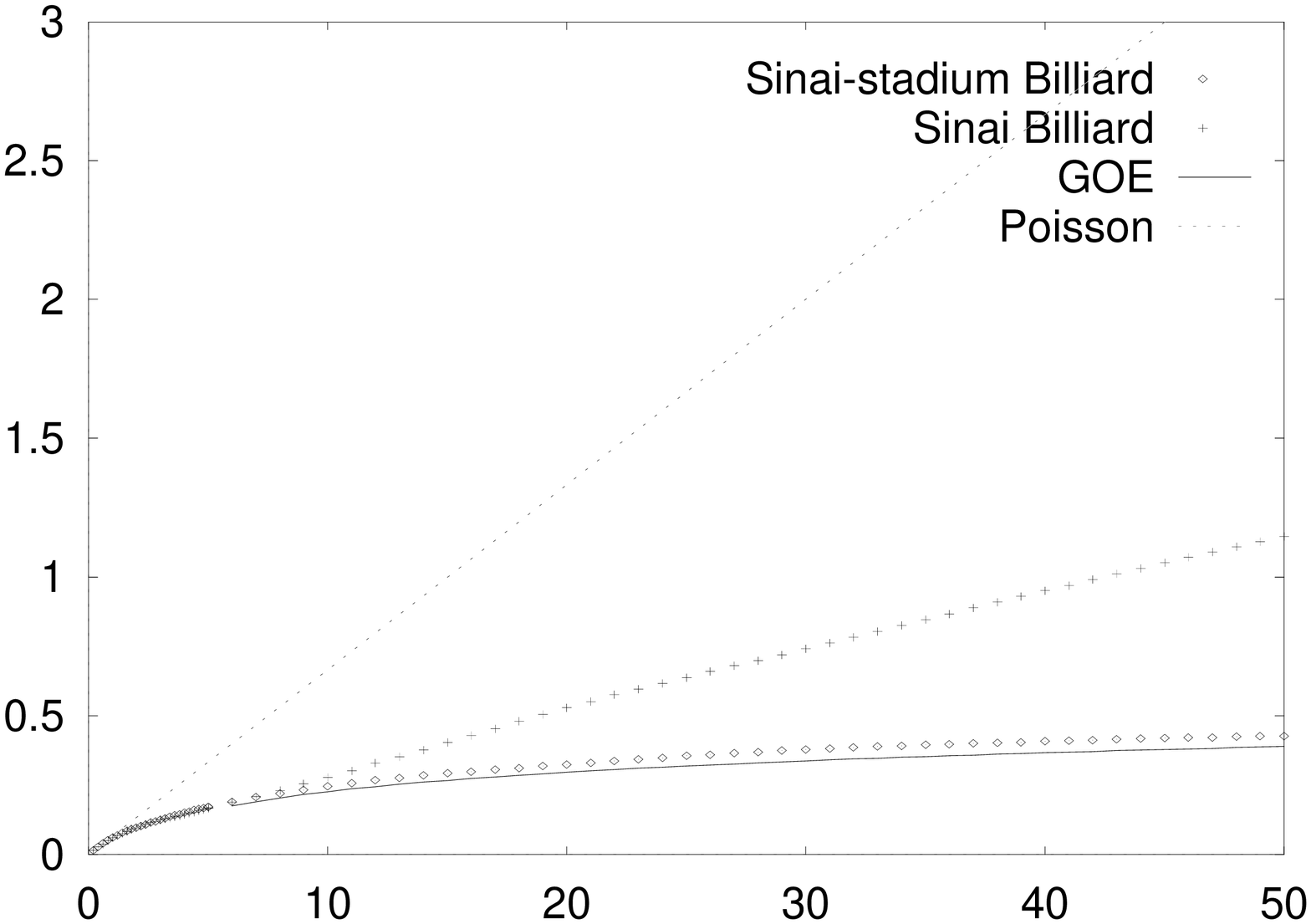,width=3.5in}}
\vskip 2mm
\caption{(Top)Experimental microwave transmission function from a Sinai billiard cavity,
(middle) Spacing statistics $P(s)$, (bottom) Spectral rigidity $\Delta_3(L)$. }
\label{fig1}
\end{figure}

\begin{figure}[htbp]
\center{
\epsfig{figure=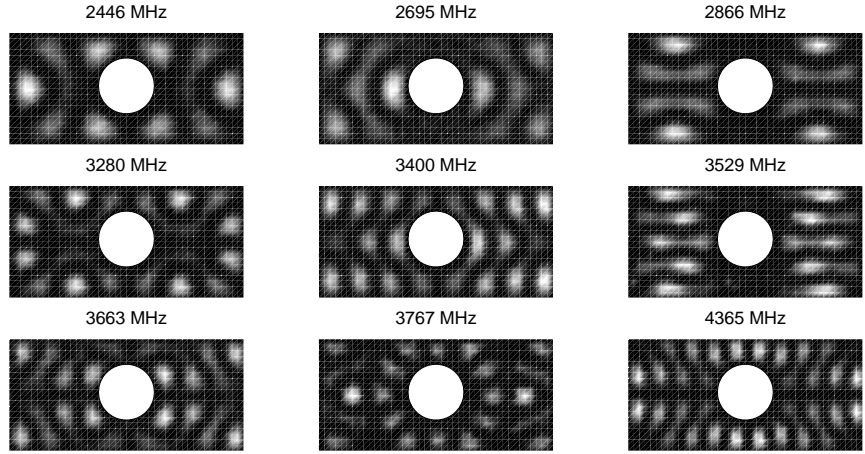,width=4.5in}}
\vskip 4mm
\caption{Representative experimental eigenfunctions of a Sinai billiard microwave cavity.
The labels represent corresponding eigenfrequencies. }
\label{fig2}
\end{figure}

\begin{figure}[htbp]
\center{
\epsfig{figure=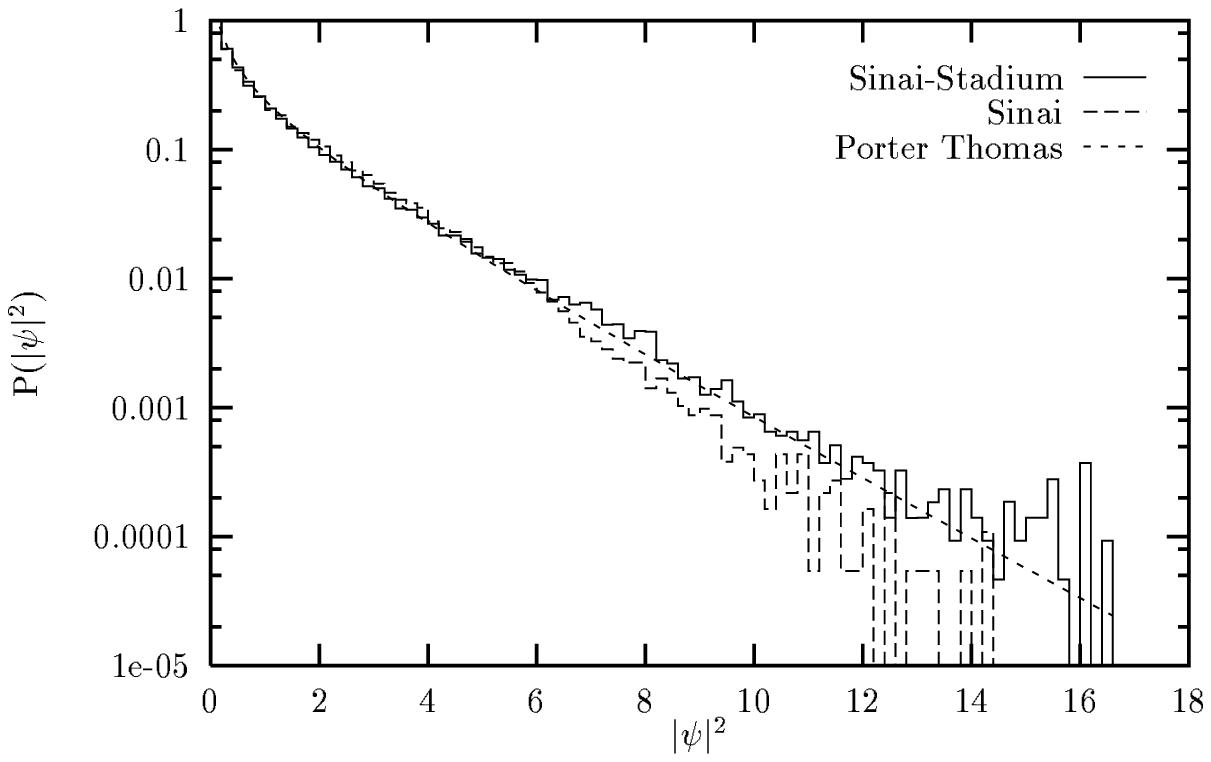,height=3.8in,angle=90}}
\epsfig{figure=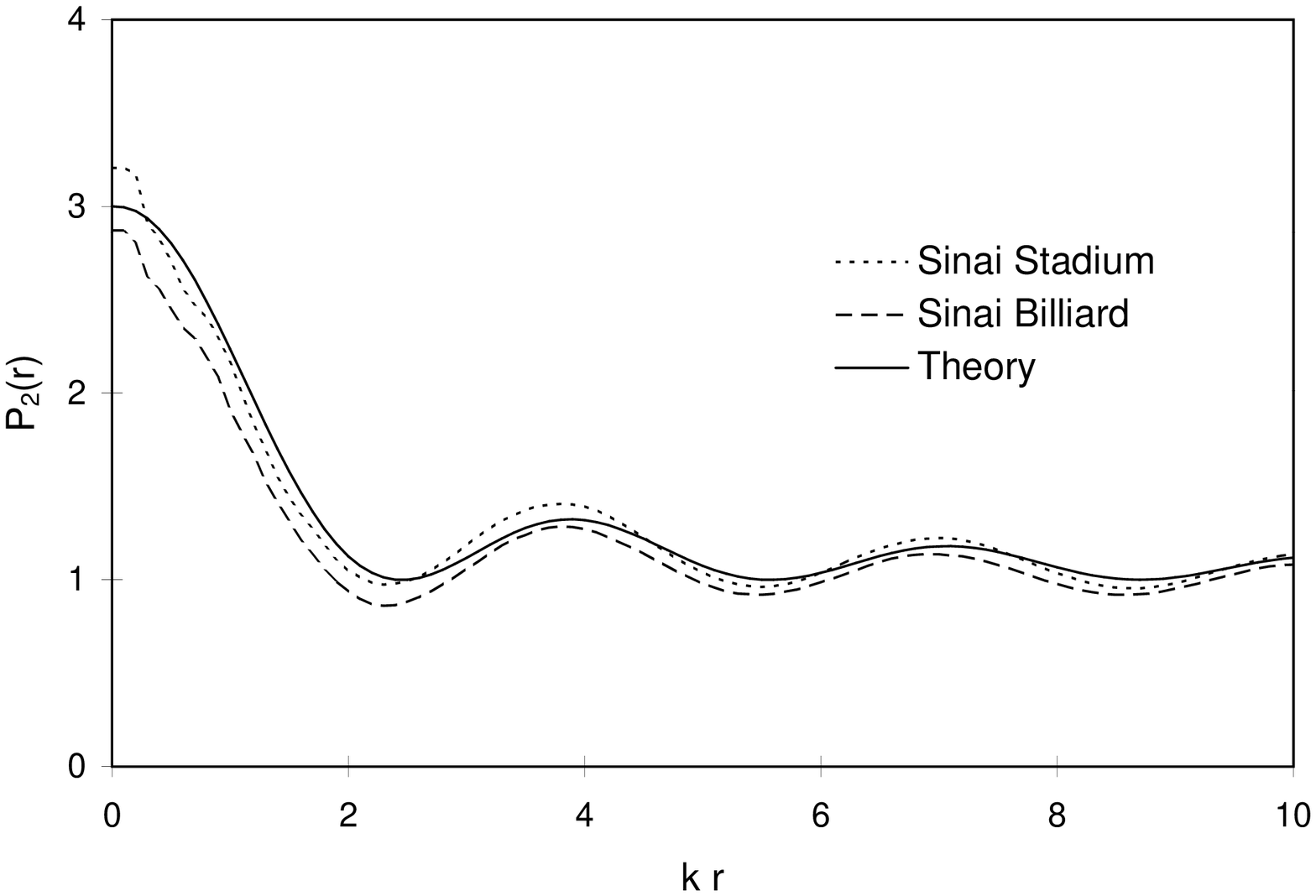,width=3.5in}
\vskip 2mm
\caption{Statistical properties of eigenfunctions of the Sinai billiard.
(top) Density Distribution $P(|\Psi |^2 )$,
(bottom) Spatial correlations. }
\label{fig3}
\end{figure}

\begin{figure}[htbp]
\center{
\epsfig{figure=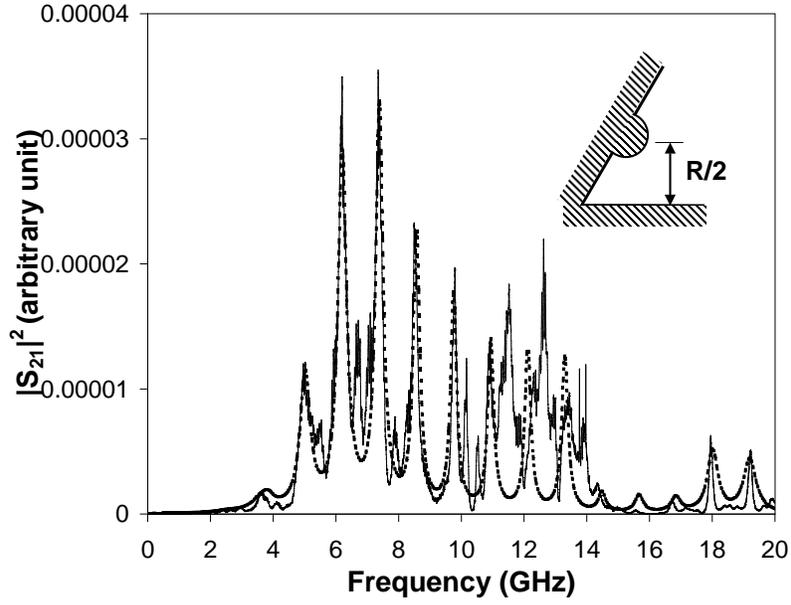,width=4.2in}}
\vskip 2mm
\caption{Quantum resonances of the 3-disk system in the fundamental domain 
with the distance between the center of the disk of radius 5cm  to the corner 
being 20cm. The dashed line uses the semiclassical resonances calculated from
the semiclassical Ruelle zeta-function.}
\end{figure}

\begin{figure}[htbp]
\center{
\epsfig{figure=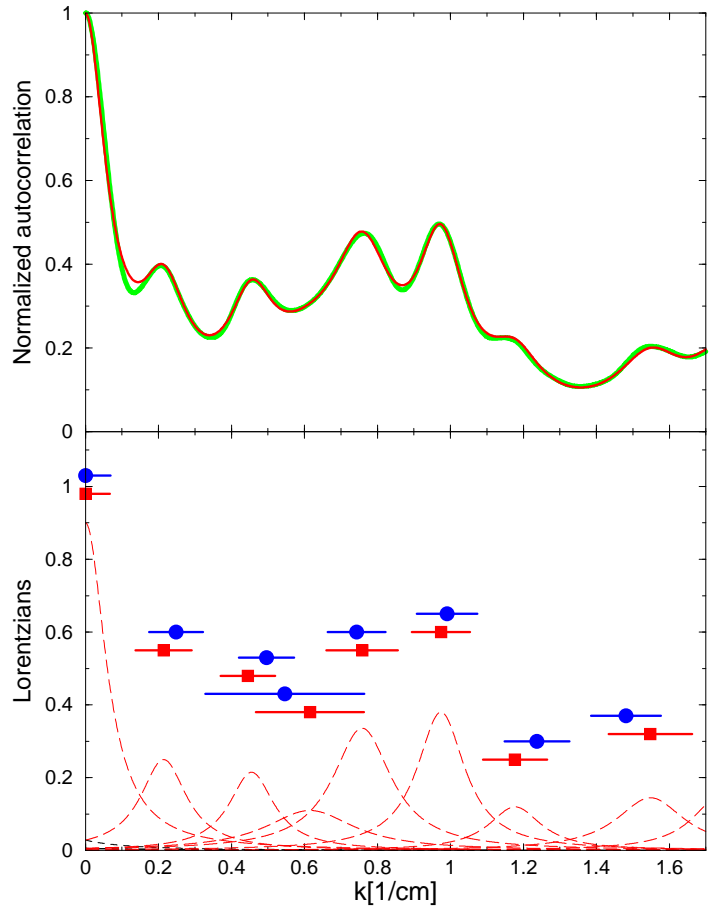,width=3.8in}}
\vskip 2mm
\caption{R-P resonances of the 3-disk system. (top) The experimental
autocorrelation $C(\kappa)$ (solid line) and the fitting (gray line).
(bottom) Dashed line is the decomposition of $C(\kappa)$ into Lorentzians.
Squares with bars are the experimental R-P resonances. Filled circles
with bars are R-P resonances calculated from classical Ruelle zeta-function.}
\end{figure}

\end{document}